\documentclass[aps,prd,reprint,nofootinbib]{revtex4-2}
  
\usepackage{amsmath,amssymb,bm}
\usepackage{graphicx}
\usepackage{xcolor}
\usepackage[hidelinks]{hyperref}
 
\begin{document}

\title{The Origin and Fate of Rapid Negative Modes in Coleman--De Luccia Tunneling}

\author{Yutaka Ookouchi}
\affiliation{Faculty of Arts and Science, Kyushu University, Fukuoka 819-0395, Japan}
 
\date{\today}

\begin{abstract}
Coleman--De Luccia tunneling can exhibit an infinite tower of rapidly
oscillating negative modes, complicating the semiclassical evaluation of
the path integral. We show that this tower does not represent infinitely
many bounce-specific instabilities, but instead is tied to a single
underlying gravitational conformal sector. This sector is exposed in two
different ways: by a large scalar gradient near the bubble wall and by the
turning point of a compact Euclidean geometry. The corresponding conformal
sector is already present in fluctuations about the false vacuum. Treating
the bounce and the false vacuum in the same well-defined gauge and with a
common regulator, we find that their high-frequency negative spectra
approach one another. The ultraviolet tower therefore cancels from the
relative negative-mode count, leaving a single bounce-specific negative
mode. {This finite-backreaction spectral cancellation provides a
corresponding picture of the result obtained in the small-backreaction limit.}
\end{abstract}

\maketitle

\section{Introduction}

The semiclassical description of false-vacuum decay has a remarkably simple structure.
Coleman and Callan showed that the decay rate is obtained from a Euclidean
bounce and the fluctuations around it
\cite{Coleman:1977py,CallanColeman1977}.  A central element of this
construction is the existence of a single negative fluctuation mode.
Its integration gives the imaginary part associated with the decay rate,
and in the thin-wall picture it has the intuitive interpretation of
changing the size of the nucleated bubble
\cite{Coleman:1987rm}.  The standard semiclassical picture therefore
relies not only on the existence of a bounce, but also on a very specific
structure of its fluctuation spectrum.

Coleman and De Luccia extended this framework to vacuum decay in the
presence of gravity \cite{Coleman:1980aw}.  Once gravity is
included, however, the negative-mode problem becomes much less
straightforward.  Early analyses showed that the quadratic fluctuation
problem can develop a wrong-sign kinetic sector, opening the possibility
of infinitely many negative modes \cite{LRT1985}.  Subsequent studies
developed several Lagrangian and Hamiltonian descriptions of the same
constrained fluctuation problem
\cite{TanakaSasaki1992,Garriga1994,Lavrelashvili2000,Tanaka1999,
KLT2000,GrattonTurok2001}.  These works made clear that gravitational
tunneling does not inherit the flat-space negative-mode story in a
trivial way.

A further difficulty is that different formulations can appear to give
different conditions for the onset of the problematic negative sector.
This issue was examined in a series of works on cosmological instantons
and Coleman--De Luccia fluctuations
\cite{GMST1998,DunneWang2006,Koehn2015}.  Lee and Weinberg gave a
particularly detailed Lagrangian analysis and showed how rapidly
oscillating negative modes arise when the coefficient of the reduced
radial kinetic term changes sign \cite{Weinberg2014}.  They also
distinguished modes associated with the bubble wall from those that can
appear near the maximal-radius region of a compact de Sitter instanton.
It was subsequently demonstrated numerically that the
same one-or-infinite pattern can occur for asymptotically flat
Coleman--De Luccia bounces, again correlated with a change in the sign
of the reduced kinetic term \cite{Gregory2018}.  Related aspects of the
problem have continued to be studied in more recent work
\cite{Bramberger2019}.

A more recent analysis shed light on part of the difference between these
approaches \cite{JinnoSato2021}.  In the
Hamiltonian formulation, the apparent condition for a negative kinetic
term depends on the choice of canonical fluctuation variable: canonical
transformations can mix a field with its conjugate momentum and thereby
change the form of the reduced kinetic term.  With an appropriate choice
of variable, the Hamiltonian and Lagrangian conditions can be made to
agree.  This explains why different criteria had appeared in the
literature.  It does not, however, by itself resolve the more basic
question emphasized there: how should the infinite set of rapid negative
modes be interpreted in the gravitational path integral?

{
A closely related perspective was recently developed by
Ivo~\cite{Ivo2025} in a covariant treatment of Coleman--De Luccia
instantons at small gravitational backreaction.  In that regime,
the one-loop contribution factorizes into gravitational and matter
parts, so that the common gravitational phase cancels between the
bounce and false vacuum.  Here we instead study the coupled
finite-backreaction problem directly, focusing on the spectral
origin of the rapid negative-mode tower.
}

That question is closely connected with a more general feature of
Euclidean gravity.  The Euclidean Einstein action is not bounded below
because rapidly varying conformal fluctuations can lower the action
without bound.  The standard Gaussian treatment therefore rotates the
conformal factor onto an imaginary integration contour
\cite{GibbonsHawkingPerry1978}.  For a de Sitter false vacuum, the same
conformal problem is present directly in the Euclidean sphere path
integral, where the trace part of the metric fluctuation has the familiar
wrong-sign kinetic term \cite{GibbonsPerry1978}; see also
Ref.~\cite{Muhlmann2026} for a recent review.  The presence of this
universal gravitational negative sector raises a natural possibility:
the rapid negative modes found in reduced Coleman--De Luccia
descriptions may not all represent new instabilities created by the
bounce.  Some of them may instead be another manifestation of the
gravitational sector that is already present in the false vacuum.

The purpose of this paper is to test this possibility directly.  We
return first to the radial Einstein--scalar system before the
gravitational constraint is eliminated.  In this unreduced description,
the local quadratic form remains regular precisely where the reduced
kinetic term becomes singular.  There is one positive and one negative
local direction on both sides of the crossing, and the negative
direction rotates smoothly as the background changes.  The singularity
of the reduced action therefore arises from eliminating the lapse
fluctuation rather than from the creation of a new local negative
direction.

This observation also unifies the two situations that had appeared
different in previous analyses.  Near a bubble wall, the negative
direction is a mixture of scalar and gravitational fluctuations because
the scalar profile varies rapidly.  Near the maximal-radius slice of a
de Sitter bounce, the same direction becomes purely the gravitational
scale-factor direction.  From this perspective, these are two different
mechanisms exposing the same underlying gravitational negative sector.

The local analysis, however, does not tell us whether this negative sector
is specific to the bounce or is already present in the false vacuum.  To
address this question, we compare the bounce and false vacuum in a gauge
that is regular on both backgrounds, using the same fluctuation variables,
boundary conditions, ultraviolet regulator, and local conformal
prescription.  Their negative
high-frequency spectra approach one another, both analytically and
numerically.  As the regulator is refined, the absolute number of
negative eigenvalues grows on both backgrounds, while their difference
remains one.  The remaining negative mode is slowly varying and
localized near the wall, consistent with the usual tunneling
deformation.  In this sense, the rapid tower is common to the
gravitational sector of the two saddles rather than an infinite set of
bounce-specific instabilities.

We also evaluate the relative one-loop determinant in the
$O(4)$-symmetric sector.  After the common ultraviolet gravitational
contribution is treated in the same way on the bounce and false vacuum,
the relative determinant is finite and the single remaining negative
direction supplies the tunneling phase.  The scalar-harmonic
decomposition on the $S^3$ slices is summarized in
Appendix~\ref{app:S3-harmonics}.  We label the scalar harmonics by
$\ell=0,1,2,\ldots$, with eigenvalues
$k_\ell^2=\ell(\ell+2)$.  The $\ell=0$ sector is precisely the
$O(4)$-symmetric radial sector studied in detail here, while $\ell>0$
describes non-$O(4)$-symmetric fluctuations.  This does not yet constitute the
complete four-dimensional Coleman--De Luccia prefactor:
non-$O(4)$-symmetric fluctuations, renormalization, and the global
gravitational integration cycle remain to be understood.

The remainder of this paper is organized as follows.  We first review
the $O(4)$-symmetric background, formulate the radial fluctuation problem
before eliminating the lapse, and derive both the exact reduced action
and its leading high-frequency part.  We next give a
configuration-space interpretation of the wall and turning-point
mechanisms.  We then relate the rapid sector to the Euclidean
gravitational conformal direction and formulate a common
bounce--false-vacuum comparison.  Finally, we present the numerical
spectral comparison, evaluate the relative determinant in the
$O(4)$-symmetric sector, and discuss the remaining tunneling direction
and the limitations of the present analysis.

\section{Reduced action and rapid negative mode problem}
\label{sec:reduced-problem}

We begin by reviewing the $O(4)$-symmetric Einstein--scalar system and
then formulate the radial fluctuation problem before the lapse is
eliminated. Starting from the complete local quadratic action, we derive
the exact lapse constraint and the familiar reduced $1/Q$ structure.
We then review the rapid negative-mode problem.

\subsection{$O(4)$-symmetric background}

We consider a canonical scalar field coupled to Einstein gravity.  For an
$O(4)$-symmetric Euclidean configuration we write
\begin{equation}
 ds^2={\cal N}(\xi)^2 d\xi^2+\rho(\xi)^2 d\Omega_3^2 ,
 \label{eq:prd-metric}
\end{equation}
where $\xi$ is the Euclidean radial coordinate, ${\cal N}$ is the lapse,
and $\rho$ is the radius of the three-sphere.  The scalar profile depends
only on $\xi$.  The Euclidean action is
\begin{equation}
\begin{split}
 S_E=2\pi^2\int d\xi\Bigg[
 &-3M_{\rm Pl}^2\rho
 \left(\frac{\dot\rho^2}{\cal N}+{\cal N}\right)\\
 &+\rho^3
 \left(\frac{\dot\phi^2}{2{\cal N}}+{\cal N}V(\phi)\right)
 \Bigg],
\end{split}
 \label{eq:prd-action}
\end{equation}
where a dot denotes a derivative with respect to $\xi$ and
$M_{\rm Pl}=(8\pi G)^{-1/2}$.

Varying the lapse and then choosing ${\cal N}=1$ gives the Euclidean
Hamiltonian constraint
\begin{equation}
 \dot\rho^2
 =
 1+\frac{\rho^2}{6M_{\rm Pl}^2}\dot\phi^2
 -\frac{\rho^2}{3M_{\rm Pl}^2}V(\phi).
 \label{eq:prd-constraint}
\end{equation}
It is useful to introduce
\begin{equation}
 Q\equiv
 \dot\rho^2-\frac{\rho^2}{6M_{\rm Pl}^2}\dot\phi^2 .
 \label{eq:prd-Q-def}
\end{equation}
Using Eq.~\eqref{eq:prd-constraint}, the same quantity can be written as
\begin{equation}
 Q=1-\frac{\rho^2}{3M_{\rm Pl}^2}V(\phi).
 \label{eq:prd-Q-potential}
\end{equation}
The quantity $Q$ will play two different roles below.  In the reduced
description it controls the sign of the kinetic term of the radial
fluctuation, while later we will show that it also has a direct geometrical
meaning in the gravitational configuration space.

\subsection{Radial fluctuations and a local gauge}
\label{subsec:radial-local-gauge}

After the gravitational constraint is eliminated, the $O(4)$-symmetric
scalar-type fluctuation can be described by a single radial variable. Rather than taking this
reduced result as the starting point, we derive it below directly from
the complete unreduced quadratic action.  This will identify precisely
where the familiar $1/Q$ factor enters.

We use throughout the fluctuation convention, as in~\cite{Weinberg2014},
\begin{equation}
 {\cal N}=1+A,\qquad
 \rho\to\rho(1+\Psi),\qquad
 \phi\to\phi+\Phi .
 \label{eq:prd-fluctuation-convention}
\end{equation}
To examine the
constrained system before eliminating the lapse, consider a radial
reparametrization
\begin{equation}
 \xi\to\xi+\epsilon(\xi).
 \label{eq:prd-radial-reparam}
\end{equation}
Around a background with ${\cal N}=1$, the fluctuations transform to
first order as
\begin{equation}
 A\to A-\dot\epsilon,\qquad
 \Psi\to\Psi-\frac{\dot\rho}{\rho}\epsilon,\qquad
 \Phi\to\Phi-\dot\phi\,\epsilon .
 \label{eq:prd-gauge-transform}
\end{equation}
Therefore, on a local patch where $\dot\phi\neq0$, one can choose
$\epsilon=\Phi/\dot\phi$ and impose $\Phi(\xi)\equiv0$ throughout that
patch.  

The $\Phi=0$ gauge is useful for exposing the local structure of the
scale-factor and lapse fluctuations, but it cannot be imposed on a
false-vacuum background for which $\dot\phi=0$. When we compare the
bounce and false vacuum directly, we will therefore use a different
gauge that is regular on both backgrounds.

We first keep the complete quadratic action in the local gauge $\Phi=0$.
Using the background equations of motion, the result can be written as
\begin{equation}
 \begin{aligned}
 \frac{S^{(2)}_{\Phi=0}}{2\pi^2}
 ={}&
 3M_{\rm Pl}^2\int d\xi\,\rho
 \Bigl[
 -\rho^2\dot\Psi^2+\Psi^2
 +2\rho\dot\rho\,A\dot\Psi
 \\
 &\hspace{31mm}
 +2A\Psi-QA^2
 \Bigr].
 \end{aligned}
 \label{eq:prd-unreduced-full}
\end{equation}
This is the $\Phi=0$ specialization of the $O(4)$-symmetric quadratic
Lagrangian of Lee and Weinberg~\cite{Weinberg2014}.  

The lapse fluctuation $A$ carries no radial derivative and is therefore an
auxiliary variable.  Varying the full quadratic action with respect to $A$
gives the exact linearized constraint
\begin{equation}
 QA=\rho\dot\rho\,\dot\Psi+\Psi .
 \label{eq:prd-lapse-constraint-full}
\end{equation}
For $Q\neq0$ this can be solved as
\begin{equation}
 A=\frac{\rho\dot\rho\,\dot\Psi+\Psi}{Q} .
 \label{eq:prd-lapse-solution}
\end{equation}
At $Q=0$, however, Eq.~\eqref{eq:prd-lapse-constraint-full} ceases to
determine $A$ and instead becomes the finite relation
$\rho\dot\rho\,\dot\Psi+\Psi=0$.  Thus the unreduced action itself is
regular at the crossing; what fails there is the step of solving the
constraint for $A$ by division by $Q$.

For $Q\neq0$, substituting Eq.~\eqref{eq:prd-lapse-solution} back into the
complete action gives, 
\begin{equation}
 \frac{S^{(2)}_{\rm red}}{2\pi^2}
 =
 3M_{\rm Pl}^2\int d\xi\,\rho
 \left[
 -\rho^2\dot\Psi^2+\Psi^2
 +\frac{\left(\rho\dot\rho\,\dot\Psi+\Psi\right)^2}{Q}
 \right].
 \label{eq:prd-reduced-exact}
\end{equation}
Expanding the square and integrating the $\Psi\dot\Psi$ term by parts,
this becomes, up to a boundary term,
\begin{equation}
 \frac{S^{(2)}_{\rm red}}{2\pi^2}
 =
 \int d\xi\,
 \left[
 \frac{\rho^5\dot\phi^2}{2Q}\dot\Psi^2
 +{\cal U}_{\Psi}(\xi)\Psi^2
 \right],
 \label{eq:prd-reduced-derived}
\end{equation}
where
\begin{equation}
 {\cal U}_{\Psi}
 =
 3M_{\rm Pl}^2
 \left[
 \rho\left(1+\frac{1}{Q}\right)
 -\frac{d}{d\xi}\left(\frac{\rho^2\dot\rho}{Q}\right)
 \right].
 \label{eq:prd-reduced-potential-Psi}
\end{equation}
The lower-derivative terms in the full unreduced action therefore
modify the reduced potential term, but they do not change the coefficient
of the highest-derivative term.  

\subsection{Leading high-frequency part}
\label{subsec:leading-high-frequency}

The exact reduction above already displays the standard $1/Q$
structure of the CDL radial fluctuation problem. Focusing on the
kinetic term in Eq.~\eqref{eq:prd-reduced-derived}, we have
\begin{equation}
 \frac{S^{(2)}_{\rm red,kin}}{2\pi^2}
 =
 \int d\xi\,
 \frac{\rho^5\dot\phi^2}{2Q}\dot\Psi^2 .
 \label{eq:prd-local-reduced}
\end{equation}
This immediately reproduces the standard rapid-mode argument. Suppose
that $Q<0$ on a finite interval and take a fluctuation supported inside
that interval whose wavelength is short compared with the background
variation scale.  For example,
\begin{equation}
 \Psi_k(\xi)=f(\xi)\sin(k\xi),
 \qquad
 \dot\Psi_k=O(k\Psi_k),
 \label{eq:prd-Psi-highk}
\end{equation}
with a slowly varying amplitude $f$.  The two-derivative part of the reduced action
then scales as
\begin{equation}
 S^{(2)}_{\rm red}[\Psi_k]
 \simeq -Ck^2+O(k^0),
 \qquad C>0 ,
 \label{eq:prd-rapid-scaling}
\end{equation}
so a negative-$Q$ interval produces an arbitrarily large set of rapidly
oscillating negative directions in the reduced description
\cite{Weinberg2014,Gregory2018}.  This is distinct from the ordinary
tunneling negative mode, which is a low-frequency property of the full
operator and can exist even when the kinetic coefficient is positive
everywhere.

The question is therefore what the same high-frequency limit looks like
before the lapse is eliminated.  The exact constraint
Eq.~\eqref{eq:prd-lapse-solution} gives, at a fixed point with $Q\neq0$
and in a generic region with $\dot\rho\neq0$,
\begin{equation}
 A_k
 =\frac{\rho\dot\rho}{Q}\dot\Psi_k
 +\frac{1}{Q}\Psi_k
 =O(k\Psi_k).
 \label{eq:prd-A-highk}
\end{equation}
Although
$A$ carries no radial derivative, it is an auxiliary variable whose
constraint makes it of the same high-frequency order as $\dot\Psi$.
Consequently,
\begin{equation}
\dot\Psi^2,\ A\dot\Psi,\ A^2=O(k^2\Psi^2)\ ,\  A\Psi=O(k\Psi^2)\ ,\ \Psi^2=O(\Psi^2).\nonumber
 \label{eq:prd-highk-counting}
\end{equation}
Thus the $A^2$ term must be retained together with
$\dot\Psi^2$ and $A\dot\Psi$, whereas the $A\Psi$ and $\Psi^2$ terms are
subleading in the high-frequency expansion.  The resulting
$(\dot\Psi,A)$ block is therefore the leading quadratic form relevant
for the rapid modes. Keeping precisely the leading $O(k^2)$ terms therefore gives
\begin{equation}
 \frac{S^{(2)}_{\rm prin}}{2\pi^2}
 =
 3M_{\rm Pl}^2\int d\xi\,\rho
 \left[
 -\rho^2\dot\Psi^2
 +2\rho\dot\rho\,A\dot\Psi
 -QA^2
 \right].
 \label{eq:prd-unreduced-local}
\end{equation}
At a fixed value of $\xi$, it is convenient to write this leading high-frequency quadratic form
in terms of $(\rho\dot\Psi,A)$:
\begin{equation}
 \frac12
 \begin{pmatrix}
  \rho\dot\Psi & A
 \end{pmatrix}
 {\cal M}
 \begin{pmatrix}
  \rho\dot\Psi\\ A
 \end{pmatrix},
 \qquad
 {\cal M}
 =
 6M_{\rm Pl}^2\rho
 \begin{pmatrix}
  -1&\dot\rho\\
  \dot\rho&-Q
 \end{pmatrix}.
 \label{eq:prd-local-matrix}
\end{equation}
Its determinant is
\begin{equation}
 \det{\cal M}
 =
 -6M_{\rm Pl}^2\rho^4\dot\phi^2 .
 \label{eq:prd-local-det}
\end{equation}
Thus, wherever $\dot\phi\neq0$, the determinant is strictly negative.
The two eigenvalues of the leading quadratic form therefore have
opposite signs, independently of whether $Q$ is positive or negative. In particular, nothing singular happens to this leading high-frequency part of the unreduced action
when $Q=0$.  At the crossing, Eq.~\eqref{eq:prd-local-det} remains finite and nonzero as long as
$\dot\phi\neq0$.  There is still one positive and one negative local
direction.  Since the matrix elements vary smoothly
with the background and the two eigenvalues remain separated, the
corresponding directions also change smoothly through $Q=0$.  The
crossing does not create a new negative direction.

\section{Geometric interpretation}
\label{sec:configuration-space}

The previous section showed that the reduced $1/Q$ behavior is generated
when the lapse fluctuation is eliminated from an otherwise regular local
two-variable problem.  We now give a geometrical interpretation of the
same result.  The main observation is that $Q$ measures the norm of the
direction normal to the bounce trajectory in the two-dimensional
configuration space spanned by the scale factor and the scalar field.

\subsection{Configuration-space geometry and the gauge-invariant radial fluctuation}

Let $q^A=(\rho,\phi)$ denote the two $O(4)$-symmetric configuration
variables.  The kinetic
part of the action \eqref{eq:prd-action} can be written as
\begin{equation}
 S_{\rm kin}
 =
 2\pi^2\int d\xi\,
 \frac{1}{2{\cal N}}\,
 G_{AB}(q)\,\dot q^A\dot q^B ,
 \label{eq:prd-config-kinetic}
\end{equation}
with
\begin{equation}
 G_{AB}
 =
 \begin{pmatrix}
 -6M_{\rm Pl}^2\rho&0\\
 0&\rho^3
 \end{pmatrix}.
 \label{eq:prd-config-metric}
\end{equation}

The bounce background defines a trajectory in this configuration space,
with tangent vector
\begin{equation}
 T^A=\dot q^A=(\dot\rho,\dot\phi).
 \label{eq:prd-tangent}
\end{equation}
Its norm is
\begin{equation}
 G(T,T)
 =
 -6M_{\rm Pl}^2\rho\,Q .
 \label{eq:prd-tangent-norm}
\end{equation}
A convenient vector orthogonal to $T^A$ is
\begin{equation}
 N^A
 =
 \left(
 \frac{\rho^2\dot\phi}{6M_{\rm Pl}^2},
 \dot\rho
 \right),
 \qquad
 G(T,N)=0 .
 \label{eq:prd-normal}
\end{equation}
Its norm is
\begin{equation}
 G(N,N)=\rho^3 Q .
 \label{eq:prd-normal-norm}
\end{equation}
Equation~\eqref{eq:prd-normal-norm} gives a direct geometrical meaning to
$Q$.  For $Q>0$, the direction normal to the background trajectory has
positive norm.  For $Q<0$, the same normal direction has negative norm.
At $Q=0$, both $T^A$ and $N^A$ have zero norm.  What becomes degenerate
there is the decomposition into tangent and normal directions; the
configuration-space metric \eqref{eq:prd-config-metric} itself remains
regular and continues to contain one negative and one positive direction.

This geometrical picture is the counterpart of the result obtained in
Sec.~\ref{subsec:leading-high-frequency}.  In the local unreduced quadratic form, the
negative direction changes smoothly as $Q$ crosses zero.  As $Q$ crosses
zero, the tangent and normal directions exchange their norm signs:
$T^A$ becomes positive-norm, while $N^A$ becomes negative-norm.

\subsection{Wall-gradient mechanism}

We now apply this picture to the two situations in which rapid negative
modes are known to appear \cite{Weinberg2014}.  First consider a bounce in which the scalar
field changes rapidly across the bubble wall.  From
Eq.~\eqref{eq:prd-Q-def}, the condition
\begin{equation}
 \frac{\rho^2\dot\phi^2}{6M_{\rm Pl}^2}
 >
 \dot\rho^2
 \label{eq:prd-wall-condition}
\end{equation}
implies $Q<0$.  In the reduced description, the coefficient of the
kinetic term then changes sign and the wall can support the rapidly
oscillating negative modes.

The normal vector in this region is
\begin{equation}
 N^A
 =
 \left(
 \frac{\rho^2\dot\phi}{6M_{\rm Pl}^2},
 \dot\rho
 \right).
 \label{eq:prd-wall-normal}
\end{equation}
Since $\dot\rho$ is generally nonzero at the wall, this direction contains
both the scale-factor and scalar components.  It is therefore not a pure
gravitational direction.  The norm is
\begin{equation}
 G(N,N)=\rho^3Q<0 .
 \label{eq:prd-wall-negative-norm}
\end{equation}

\subsection{Turning-point mechanism and common interpretation}

The second case occurs in a de Sitter CDL bounce.  The Euclidean scale
factor increases from one pole, reaches a maximum at some
$\xi=\xi_*$, and then decreases.  At the maximal-radius slice,
$\dot\rho(\xi_*)=0$.  If the scalar field is still evolving there,
Eq.~\eqref{eq:prd-Q-def} immediately gives
\begin{equation}
 Q_*
 =
 -\frac{\rho_*^2\dot\phi_*^2}{6M_{\rm Pl}^2}
 <0 .
 \label{eq:prd-Qstar}
\end{equation}
At this point,
\begin{equation}
 N_*^A
 =
 \left(
 \frac{\rho_*^2\dot\phi_*}{6M_{\rm Pl}^2},
 0
 \right).
 \label{eq:prd-Nstar}
\end{equation}
The normal direction is therefore purely along the scale-factor
coordinate.  Its norm is
\begin{equation}
 G(N,N)_*
 =
 -\frac{\rho_*^5\dot\phi_*^2}{6M_{\rm Pl}^2}
 <0 .
 \label{eq:prd-Nstar-norm}
\end{equation}
Thus the turning point forces the normal direction to have negative norm,
independently of the detailed shape of the bubble wall.  The underlying
gravitational character of the negative direction is especially
transparent here.

\section{The gravitational conformal sector and a common
bounce--false-vacuum formulation}
\label{sec:common-formulation}

The analysis of the previous section reveals a gravitational negative
direction underlying both the wall and turning-point rapid sectors.
The remaining question is whether the associated rapid negative modes
represent instabilities specific to the bounce.  This cannot be decided
from the local structure of the bounce alone: the relevant comparison is
with the corresponding false vacuum, which itself contains the familiar
conformal negative sector of Euclidean gravity.

We therefore first review the Euclidean conformal-factor problem and then
formulate the bounce and false-vacuum fluctuation problems in a common
gauge.  This will allow us to compare their high-frequency negative
spectra directly.

\subsection{The Euclidean conformal direction}

Consider the Euclidean Einstein action
\begin{equation}
 S_E^{\rm grav}[g]
 =
 -\frac{M_{\rm Pl}^2}{2}
 \int d^4x\,\sqrt{g}\,R
 +\text{boundary terms}.
 \label{eq:prd-EH-action}
\end{equation}
For a positive-definite metric, perform the conformal transformation
$\widetilde g_{ab}=\Omega^2 g_{ab}$.  In four dimensions,
\begin{equation}
 \widetilde R
 =
 \Omega^{-2}R-6\Omega^{-3}\Box\Omega .
 \label{eq:prd-conformal-R}
\end{equation}
After integration by parts, with the boundary contribution absent or fixed by the boundary conditions,
\begin{equation}
\begin{split}
 \widetilde S_E^{\rm grav}
 &=
 -\frac{M_{\rm Pl}^2}{2}
 \int d^4x\,\sqrt{g}\,
 \left[
 \Omega^2R+6(\nabla\Omega)^2
 \right]
 \\
 &\quad
 +\text{boundary terms}.
\end{split}
 \label{eq:prd-conformal-action}
\end{equation}
The overall sign in Eq.~\eqref{eq:prd-conformal-action} shows that a
rapidly varying conformal factor can lower the Euclidean gravitational
action without bound.  This is the conformal-factor problem emphasized
by Gibbons, Hawking, and Perry \cite{GibbonsHawkingPerry1978}.  It is not
a property created by a particular bounce; it is already present in the
Euclidean gravitational path integral.

For the de Sitter false vacuum this statement can be made especially
explicit.  Its Euclidean continuation is the round $S^4$, and in a
covariant gauge the trace part of the metric fluctuation has the
wrong-sign kinetic term characteristic of the conformal mode problem
\cite{GibbonsPerry1978}; see also Ref.~\cite{Muhlmann2026} for a
recent review.  A careful contour prescription leaves a finite residual
phase associated with the low $S^4$ trace harmonics; in four dimensions
these are the $\ell=0$ and $\ell=1$ conformal modes that determine the
residual phase of the $S^4$ gravitational path integral
\cite{Polchinski1989}.  Thus the
false-vacuum sphere already contains the conformal sector whose
high-frequency part will be compared with the bounce below.

{
This relative viewpoint has also been implemented directly in the
Coleman--De Luccia path integral by Ivo~\cite{Ivo2025}.
In the small-backreaction limit, the common gravitational
contribution, including its phase, cancels between the bounce and
false vacuum, leaving the usual tunneling phase.
}
\footnote{The
$S^4$ harmonic label $\ell$ in this statement should not be identified
mode by mode with the $S^3$ partial-wave label used for the
Coleman--De Luccia fluctuations in this paper.}

In our Einstein--scalar problem, the radial negative direction near the
bubble wall is not, mode by mode, identical to the pure-gravity trace
fluctuation considered in Ref.~\cite{GibbonsHawkingPerry1978}, because
scalar and metric fluctuations are mixed there.  By contrast, at the
turning point, its gravitational character is manifest, since the normal
direction becomes purely the scale-factor direction.  The comparison of
the bounce and false vacuum with a common gauge choice below shows that the rapid
negative sectors associated with the wall and the turning point are not
independent ultraviolet instabilities, but approach the same gravitational
negative sector at high frequency.

A negative Euclidean Gaussian is not integrated along the real direction
in the standard conformal-factor prescription.  If a local negative
eigenmode $x_-$ contributes
\begin{equation}
 S_-^{(2)}=\frac{\lambda_-}{2}x_-^2,
 \qquad \lambda_-<0,
 \label{eq:prd-negative-gaussian}
\end{equation}
one rotates the integration variable according to
\begin{equation}
 x_-=iy_- .
 \label{eq:prd-conformal-rotation}
\end{equation}
The quadratic action then becomes
\begin{equation}
 S_-^{(2)}
 =
 \frac{|\lambda_-|}{2}y_-^2>0 ,
 \label{eq:prd-rotated-gaussian}
\end{equation}
and the local Gaussian integral converges.  

The vacuum decay rate depends on the bounce contribution relative to the
false-vacuum contribution, not on the bounce determinant by itself.  The
two fluctuation problems must therefore be defined using the same
variables, gauge condition, boundary conditions, regulator, functional
measure, and contour prescription.  As we will show below, the leading negative high-frequency contribution
is common to the two backgrounds.

Suppose the bounce and false vacuum contain $n_-^{\rm B}$ and
$n_-^{\rm FV}$ negative directions, respectively, defined with the same
prescription.  Rotating each common negative Gaussian in the same way
gives a relative phase that depends only on the difference
\begin{equation}
 \Delta n_-
 =
 n_-^{\rm B}-n_-^{\rm FV}.
 \label{eq:prd-relative-count}
\end{equation}
The rapidly increasing number of ultraviolet gravitational directions can
therefore cancel from the relative phase even when the absolute numbers
on the two backgrounds are large.  The ultraviolet analysis alone does
not determine the finite value of $\Delta n_-$: any residual mismatch is
an infrared property of the two fluctuation problems and must be found by
a matched spectral comparison.

{
This relative cancellation has an important precedent in the
small-backreaction analysis of Ref.~\cite{Ivo2025}.  There the
one-loop path integral factorizes, to leading order in the
gravitational backreaction, as
\begin{equation}
 Z_{\rm CdL}
 =
 Z_{\rm GR}(S^4)\,
 Z_{\phi}(\mathrm{bounce})
 [1+O(G_N)],
 \label{eq:prd-ivo-factorization}
\end{equation}
with an analogous factorization for the false vacuum.  The common
gravitational factor, including its phase, therefore cancels in
the ratio defining the decay rate, leaving the usual tunneling
phase.

At finite backreaction, however, the scalar and gravitational
fluctuations are coupled and such a factorization need not hold.
Ref.~\cite{Ivo2025} proposes to follow the phase in this regime
by tracking eigenvalue crossings.  Here we instead examine
directly whether the rapid negative modes of the coupled
Coleman--De Luccia problem can be matched to the negative
gravitational sector of the false vacuum, without assuming a
gravity--matter factorization.}

\subsection{Common gauge choice and high-frequency comparison}

For a direct comparison, we use the same
fluctuation variables introduced in Eq.~\eqref{eq:prd-fluctuation-convention},
with the radial reparametrization law given in
Eq.~\eqref{eq:prd-gauge-transform}.  The local choice $\Phi=0$ used in Sec.~\ref{subsec:radial-local-gauge}
requires $\dot\phi\neq0$ and therefore cannot be used on the false vacuum,
where the scalar is constant.  Choosing $\dot\epsilon=A$ sets $A=0$.
This condition does not depend on
the value of $\dot\phi$ and can therefore be imposed on both the bounce
and the false vacuum. 

For the ultraviolet behavior we only need the terms with two
derivatives on the fluctuations, together with mixing terms that remain
at the same order in the short-wavelength limit.  In the $A=0$ gauge,
the gravitational kinetic term contributes
$
 -3M_{\rm Pl}^2\rho^3\dot\Psi^2$, while the scalar kinetic term contributes, to quadratic order,
$
 \rho^3\dot\Phi^2/2
 +3\rho^3\dot\phi\,\Psi\dot\Phi .
$
The mixing term may be integrated by parts.  Keeping only the part that
remains at the same order in the high-frequency limit gives
$ 3\rho^3\dot\phi\,\Psi\dot\Phi
 \simeq
 -3\rho^3\dot\phi\,\Phi\dot\Psi ,
$
where terms containing derivatives of the background are lower order in
the local short-wavelength expansion.  Combining these terms gives
\begin{equation}
 \frac{S^{(2)}_{\rm kin}}{2\pi^2}
 =
 \int d\xi
 \left[
 -3M_{\rm Pl}^2\rho^3\dot\Psi^2
 +\frac{\rho^3}{2}\dot\Phi^2
 -3\rho^3\dot\phi\,\Phi\dot\Psi
 \right]
 +\cdots ,
 \label{eq:prd-common-derivative-action}
\end{equation}
where the omitted terms contain fewer derivatives on the fluctuations.
The first term is the radial scale-factor contribution and carries the
negative sign characteristic of Euclidean gravity.

It is useful to remove the positive background normalization by defining
\begin{equation}
 g=\sqrt{6}\,M_{\rm Pl}\rho^{3/2}\Psi,\qquad
 s=\rho^{3/2}\Phi ,
 \label{eq:prd-canonical-gs}
\end{equation}
and
\begin{equation}
 c=\sqrt{\frac32}\frac{\dot\phi}{M_{\rm Pl}} .
 \label{eq:prd-c-def}
\end{equation}
Before taking the local high-frequency limit, the derivative structure in
these variables can be written as
\begin{equation}
 \frac{S_{\rm der}^{(2)}}{2\pi^2}
 =
 \int d\xi
 \left[
 -\frac12({\cal D}g)^2
 +\frac12({\cal D}s)^2
 -c\,s\,{\cal D}g
 \right],
 \label{eq:prd-D-action}
\end{equation}
where
\begin{equation}
 {\cal D}
 =
 \frac{d}{d\xi}
 -\frac32\frac{\dot\rho}{\rho}.
 \label{eq:prd-D-def}
\end{equation}
At wavelengths much shorter than the scale over which the background
changes, the background coefficients may be treated locally as constants
and ${\cal D}$ may be replaced by $d/d\xi$ at leading order.  The
derivative part of the quadratic action then becomes
\begin{equation}
 \frac{S^{(2)}_{\rm kin}}{2\pi^2}
 \simeq
 \int d\xi
 \left[
 -\frac12\dot g^2+\frac12\dot s^2-c\,s\dot g
 \right].
 \label{eq:prd-high-k-action}
\end{equation}

For a local Fourier mode proportional to $e^{ik\xi}$, the bounce
quadratic form associated with Eq.~\eqref{eq:prd-high-k-action} is
\begin{equation}
 H_{\rm B}(k)
 =
 \begin{pmatrix}
 -k^2&ick\\
 -ick&k^2
 \end{pmatrix}.
 \label{eq:prd-HB}
\end{equation}
Its eigenvalues are
\begin{equation}
 \lambda_\pm^{\rm B}(k)
 =
 \pm |k|\sqrt{k^2+c^2}.
 \label{eq:prd-lambdaB}
\end{equation}

The appearance of both positive and negative high-frequency eigenvalues
does not conflict with the conformal-factor analysis on the de Sitter
sphere.  The latter concerns the gravitational trace sector alone,
whereas the present operator acts on the coupled gravitational--scalar
system.  In the limit of vanishing scalar--gravity mixing, the two
branches reduce to the negative gravitational conformal sector and the
positive scalar sector, respectively.  Thus it is the negative branch
$\lambda_-$, rather than the full $\lambda_\pm$ spectrum, that should be
compared with the conformal sector of the false vacuum.

Thus the coupled scalar--gravity system has one positive and one negative
high-frequency eigenvalue.

On the false vacuum, $\dot{\phi}=0$ and hence $c=0$.  The mixing disappears,
\begin{equation}
 H_{\rm FV}(k)
 =
 \begin{pmatrix}
 -k^2&0\\
 0&k^2
 \end{pmatrix},
 \qquad
 \lambda_\pm^{\rm FV}(k)=\pm k^2 .
 \label{eq:prd-HFV}
\end{equation}
The important point is that the false vacuum already contains a negative
high-frequency gravitational sector.  For the bounce,
\begin{equation}
 \lambda_-^{\rm B}(k)
 =
 -k^2-\frac{c^2}{2}+O(k^{-2}),
 \label{eq:prd-lambdaB-asymp}
\end{equation}
and therefore
\begin{equation}
 \frac{\lambda_-^{\rm B}(k)}
 {\lambda_-^{\rm FV}(k)}
 =
 1+\frac{c^2}{2k^2}+O(k^{-4})
 \to1
 \qquad (|k|\to\infty).
 \label{eq:prd-UV-ratio}
\end{equation}
This is the central ultraviolet result.  The mixed scalar--gravity
negative modes on the bounce do not introduce a new high-frequency
instability.  As the wavelength is shortened, their negative eigenvalues
approach those of the gravitational conformal sector already present in
the false vacuum.

\section{Numerical comparison of the bounce and false vacuum}
\label{sec:numerical-comparison}

The analysis of Sec.~\ref{sec:common-formulation} shows that the negative
high-frequency eigenvalues of the bounce approach those of the false
vacuum.  We now test this statement directly on numerical CDL
backgrounds.  The essential point is to compare the two fluctuation
problems with the same variables, gauge condition, boundary conditions, and
discretization rule.  

We first study a wall-induced example with a Minkowski false vacuum and
then repeat the comparison for a de Sitter CDL bounce, where the negative
region is forced by the turning point of the Euclidean scale factor.

\subsection{Minkowski wall example and spectral comparison}

Following the asymptotically flat wall example studied in~\cite{Gregory2018}, we use the quartic potential
\begin{equation}
 V(\phi)
 =
 \frac{\lambda_q}{4}\phi^4
 -\frac{\lambda_q}{3}(\phi_m+\phi_t)\phi^3
 +\frac{\lambda_q}{2}\phi_m\phi_t\phi^2 ,
 \label{eq:prd-quartic-potential}
\end{equation}
with
\begin{equation}
 \lambda_q=128,\qquad
 \phi_t=0.69\,M_{\rm Pl},\qquad
 \phi_m=0.276\,M_{\rm Pl}.
 \label{eq:prd-quartic-parameters}
\end{equation}
The false vacuum is at $\phi=0$ and has vanishing vacuum energy.  Its
Euclidean geometry is therefore
\begin{equation}
 \phi_{\rm FV}=0,\qquad
 \rho_{\rm FV}(\xi)=\xi .
 \label{eq:prd-Minkowski-FV}
\end{equation}

Solving the $O(4)$-symmetric background equations numerically gives a CDL
bounce for which $Q$ becomes negative in the wall region.  For the
solution used here, the minimum value of $Q$ is $Q_{\rm min}\simeq -0.0862$.  Thus the reduced action
contains the wall-localized rapid negative modes
described in Sec.~\ref{sec:reduced-problem}.

For the spectral comparison, however, we do not use the singular reduced
description.  Unlike the local high-frequency analysis of
Sec.~\ref{sec:common-formulation}, the numerical calculation retains the
full position dependence of the bounce background and all lower-derivative
terms in the quadratic fluctuation operator.  We work in the common
$A=0$ gauge of Sec.~\ref{sec:common-formulation}.  The two
quadratic forms are constructed with the same radial regulator.  In
particular, the same Dirichlet endpoint conditions are imposed on the
canonical variables $g$ and $s$, and the same radial discretization is
used on both backgrounds. 
 
\begin{figure}[t]
 \centering
 \includegraphics[width=0.96\columnwidth]{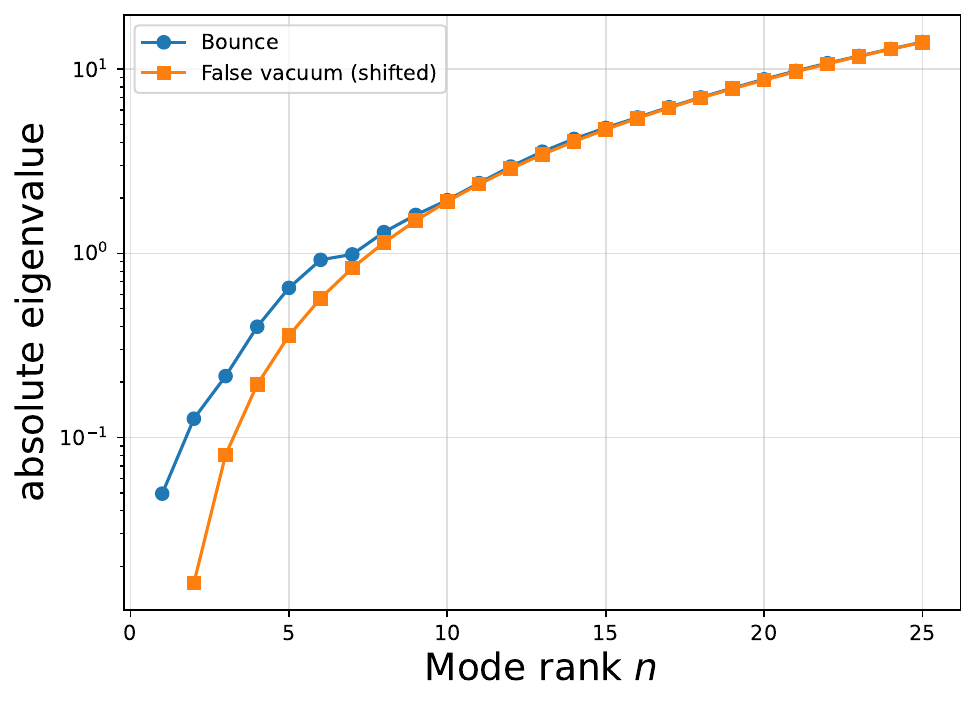}
 \caption{Negative spectrum in the decay of a metastable Minkowski vacuum in
 the common $A=0$ gauge, using the radial cutoff
 $L_{\rm B}=L_{\rm FV}=20$ and $N_{\rm grid}=8000$.  The false-vacuum mode
 number is shifted by one to display the additional low-frequency bounce
 direction.  The remaining negative eigenvalues pair increasingly closely
 toward the ultraviolet.}
 \label{fig:prd-wall-spectrum}
\end{figure}

We computed the spectrum directly from the numerical
bounce background. Since the Minkowski radial interval is noncompact, we
introduce a finite radial regulator and impose the same cutoff on the bounce
and false-vacuum backgrounds,
\begin{equation}
 L_{\rm B}=L_{\rm FV}=20,
 \label{eq:prd-wall-common-cutoff}
\end{equation}
with $N_{\rm grid}=8000$ equal radial steps. Here $L_{\rm B}$ and
$L_{\rm FV}$ denote the regulated radial endpoints. See Appendix B for calculation details. At this resolution the
regulated negative-mode
counts are
\begin{equation}
 n_-^{\rm B}=8000,\qquad
 n_-^{\rm FV}=7999,\qquad
 \Delta n_-=1 .
 \label{eq:prd-wall-negative-count}
\end{equation}
Figure~\ref{fig:prd-wall-spectrum} shows the first 25 negative-mode
magnitudes for the reference calculation with
$L_{\rm B}=L_{\rm FV}=20$ and $N_{\rm grid}=8000$.  The false-vacuum mode
number is shifted by one so that the additional low-frequency bounce
direction is displayed separately.  After this shift, the remaining
eigenvalues pair increasingly closely toward the ultraviolet.  Thus Fig.~\ref{fig:prd-wall-spectrum} exhibits two characteristic
features: a single unmatched negative mode and increasingly accurate
pairing of the remaining spectrum at high mode number.

\begin{figure}[t]
 \centering
 \includegraphics[width=0.96\columnwidth]{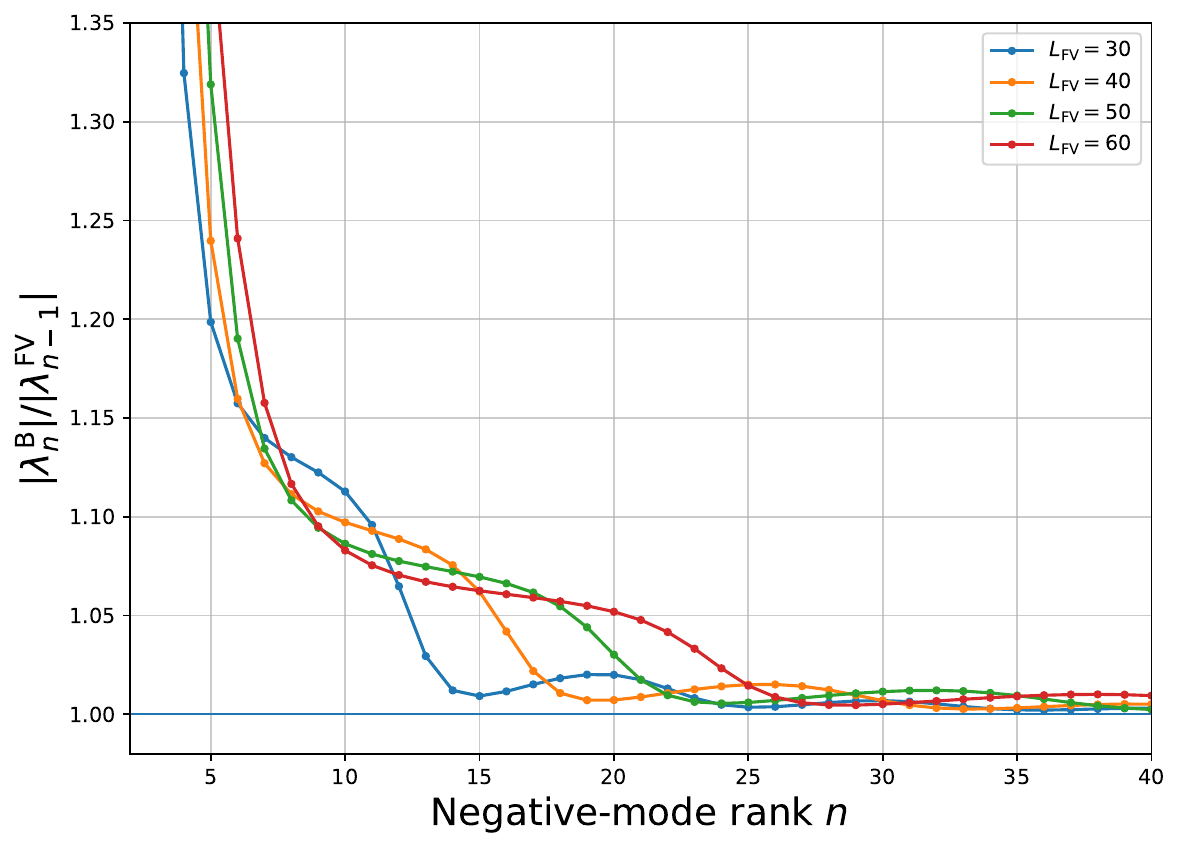}
 \caption{Dependence of the paired negative-mode spectrum on the common
 radial regulator in the decay of a metastable Minkowski vacuum.  The ratio
 $|\lambda_n^{\rm B}|/|\lambda_{n-1}^{\rm FV}|$ is shown for
 $L_{\rm FV}=30,40,50,60$, with $L_{\rm B}=L_{\rm FV}$ and fixed radial
 step $h=0.0075$. As the radial cutoff is increased, the approach
 of the paired eigenvalue ratio to unity occurs at progressively larger
 mode number.}
 \label{fig:prd-wall-cutoff-scan}
\end{figure}

The detailed behavior at low mode number, however, should not be
interpreted as universal.  To test its dependence on the finite radial
cutoff, in Fig.~\ref{fig:prd-wall-cutoff-scan} we repeat the paired
spectral comparison for several larger values of the common cutoff, while keeping the radial step fixed at
$h=0.0075$.  The corresponding grid sizes are $N_{\rm grid}=4000$,
$5333$, $6667$, and $8000$ for $L_{\rm B}=L_{\rm FV}=30$, $40$, $50$, and $60$, respectively.  As the common radial cutoff is increased, the approach of the
paired eigenvalue ratio to unity occurs at progressively larger mode
number.  This shows that the detailed low-mode-number behavior of the
ratio depends on the finite radial cutoff.

What remains robust throughout the cutoff scan is the qualitative
spectral structure: after accounting for the single unmatched negative
mode, the bounce and false-vacuum eigenvalues approach one another at
higher mode number, in accord with the analytical high-frequency result
in Eq.~\eqref{eq:prd-UV-ratio}.  We therefore do not attach physical
significance to the detailed values of the lowest regulated eigenvalues
or to the particular mode number at which the ratio begins to settle near
unity.  The robust observations are the one-unit mismatch in the
regulated negative-mode count,
\begin{equation}
 \Delta n_-=1,
\end{equation}
together with the asymptotic pairing of the remaining negative modes.

\subsection{de Sitter CDL bounce}

We next add a positive constant to the same scalar potential,
\begin{equation}
 V_{\rm dS}(\phi)=V(\phi)+V_0,
 \qquad
 V_0=0.01\,M_{\rm Pl}^4 .
 \label{eq:prd-dS-potential}
\end{equation}
The corresponding Euclidean de Sitter geometry has two regular poles,
and the scale factor reaches a maximum between them.  At the
maximal-radius slice, $\dot\rho=0$.  For the CDL solution considered here,
the scalar field is still evolving at that point, and hence
\begin{equation}
 Q
 =
 -\frac{\rho^2\dot\phi^2}{6M_{\rm Pl}^2}
 <0
 \label{eq:prd-dS-Qnegative}
\end{equation}
there.  This realizes the turning-point mechanism discussed in
Sec.~\ref{sec:configuration-space}.

For the de Sitter comparison it is convenient to introduce a common
dimensionless coordinate
\begin{equation}
 x=\frac{\xi}{L},
 \qquad
 0\leq x\leq1 ,
 \label{eq:prd-common-x}
\end{equation}
where $L_{\rm B}$ and $L_{\rm FV}$ denote the pole-to-pole proper
lengths of the bounce and false-vacuum geometries, respectively. We divide this interval into $N_{\rm grid}$ equal steps, leaving
$N_{\rm grid}-1$ interior points, and use the same midpoint
discretization on both backgrounds.  Regularity at the two poles is
implemented as Dirichlet conditions on the canonical variables $g$ and
$s$.

For the compact de Sitter solution, we integrate the numerical bounce
through the wall and sufficiently far into the false-vacuum region.  Once
the scalar field has approached the false vacuum to numerical accuracy,
the remaining portion of the geometry is continued using the analytic
false-vacuum de Sitter solution.  This construction retains the wall and
turning-point regions of the numerical bounce while providing a regular
completion to the second pole.  The same background construction is used
in both the spectral and determinant comparisons below.

\begin{figure}[t]
 \centering
 \includegraphics[width=0.94\columnwidth]{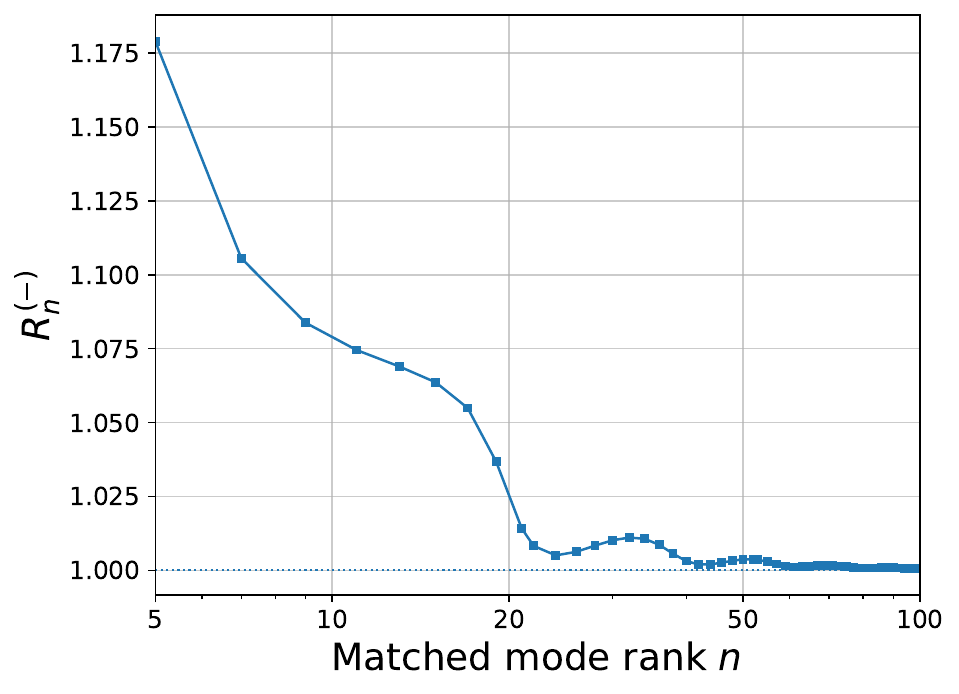}
 \caption{Length-scaled ratio of the negative eigenvalues for the compact
 de Sitter bounce and false vacuum at $N_{\rm grid}=8000$,
 $R_n^{(-)}=|\lambda_{n,{\rm B}}^{(-)}|L_{\rm B}^2/
 (|\lambda_{n,{\rm FV}}^{(-)}|L_{\rm FV}^2)$.
 The additional unmatched bounce negative mode is removed before the
 matched mode rank $n$ is assigned.  Fifty directly calculated mode ranks
 between $n=5$ and $n=100$ are shown.  The ratio approaches unity toward the ultraviolet after the
 leading difference in proper length has been removed.}
 \label{fig:prd-dS-negative-ratio}
\end{figure}

We again compare the bounce and the corresponding false vacuum. Since the two proper lengths are not
identical, a direct comparison of eigenvalues would contain a trivial
geometric mismatch.  For a second-order fluctuation operator, the
high-frequency eigenvalues scale as
$k_n^2\sim(n\pi/L)^2$.  Thus the difference between $L_B$ and
$L_{\rm FV}$ produces a leading $L^{-2}$ mismatch even when the
ultraviolet operators are otherwise identical.  We remove this geometric
scaling by comparing the dimensionless combinations
$|\lambda_n|L^2$.  Defining the ratio for the negative sector,
\begin{equation}
 R_n^{(-)}
 =
 \frac{|\lambda_{n,{\rm B}}^{(-)}|L_{\rm B}^2}
      {|\lambda_{n,{\rm FV}}^{(-)}|L_{\rm FV}^2},
 \label{eq:prd-Rpm}
\end{equation}
we recompute the full regulated bounce and false-vacuum spectra at
$N_{\rm grid}=8000$.  At this resolution the regulated counts are
$n_-^{\rm B}=7999$ and $n_-^{\rm FV}=7998$, so that one negative
direction is unmatched on the bounce.  We remove this unmatched bounce
mode before assigning the matched mode rank $n$.  Figure~\ref{fig:prd-dS-negative-ratio} shows 50
directly calculated values between $n=5$ and $n=100$. The negative eigenvalues exhibit
the same pairing seen analytically in Eq.~\eqref{eq:prd-UV-ratio}.  

The high-frequency negative gravitational sector is common to the two
backgrounds, while the regulated difference in the number of negative
directions is
\begin{equation}
 \Delta n_-
 =
 n_-^{\rm B}-n_-^{\rm FV}
 =1 .
 \label{eq:prd-dS-delta-n}
\end{equation}
Thus the relative result is the same as in the wall-induced Minkowski
example, even though the mechanism that makes $Q$ negative is different.

\section{The $\ell=0$ one-loop contribution}
\label{sec:l0-prefactor}

The numerical comparison of Sec.~\ref{sec:numerical-comparison} shows
that the growing high-frequency negative sector is common to the bounce
and false vacuum, one additional negative direction remains on the bounce.  We now evaluate its relative one-loop determinant for $O(4)$-symmetric fluctuations.

\subsection{Relative determinant in the $O(4)$-symmetric sector}

Schematically, the decay rate is
\begin{equation}
 \frac{\Gamma}{V}={\cal A}e^{-B},
 \qquad
 B=S_E[{\rm B}]-S_E[{\rm FV}] ,
 \label{eq:prd-decay-rate}
\end{equation}
where ${\cal A}$ contains fluctuation determinants, zero-mode factors,
ghost contributions, and the treatment of negative directions.

For an $O(4)$-symmetric bounce, the full one-loop contribution may be
organized in the $S^3$ scalar harmonics reviewed in
Appendix~\ref{app:S3-harmonics}, as
\begin{equation}
 \log{\cal A}
 \sim
 -\frac12
 \sum_{\ell=0}^{\infty} d_\ell
 \log
 \frac{\det{}'{\cal O}_{\rm B}^{(\ell)}}
      {\det{\cal O}_{\rm FV}^{(\ell)}}
 +\cdots ,
 \label{eq:prd-partial-wave}
\end{equation}
where $d_\ell$ is the degeneracy of the corresponding $S^3$ harmonic.
The ellipsis denotes zero-mode factors, ghost contributions, and
renormalization terms.  In this paper we focus on the $\ell=0$
contribution.

For the compact de Sitter geometries considered here, the proper radial
coordinate covers different intervals for the bounce and the false
vacuum,
\begin{equation}
 0\leq\xi\leq L_{\rm B},
 \qquad
 0\leq\xi\leq L_{\rm FV},
 \label{eq:prd-radial-intervals}
\end{equation}
respectively.  For the reference solution of
Sec.~\ref{sec:numerical-comparison},
\begin{equation}
 L_{\rm B}\simeq53.7,
 \qquad
 L_{\rm FV}\simeq54.4
 \label{eq:prd-lengths}
\end{equation}
in units with $M_{\rm Pl}=1$.  The bounce and false vacuum are evaluated
in the common $A=0$ gauge using the canonical variables
\begin{equation}
 g=\sqrt6\,M_{\rm Pl}\rho^{3/2}\Psi,
 \qquad
 s=\rho^{3/2}\Phi .
 \label{eq:prd-gs-repeat}
\end{equation}

The different pole-to-pole lengths require some care in defining the
relative determinant.  In Sec.~\ref{sec:numerical-comparison} we mapped
the spectra to a common dimensionless interval only to expose the
ultraviolet pairing after removing the trivial $L^{-2}$ scaling of the
eigenvalues.  That change of coordinate should not be confused with a
definition of the functional measure.  As emphasized in
Sec.~\ref{sec:common-formulation}, the determinant magnitude must instead
be obtained from the same regulated Gaussian integral that defines the
measure and the second-variation matrix of the action.  We therefore carry out the determinant
normalization directly in the proper radial coordinate $\xi$.

Since this point is essential for the numerical normalization of the
determinant, we give the complete finite-dimensional derivation in
Appendix~\ref{app:measure-determinant}.  Here we summarize only the
result in the proper radial coordinate.

For the midpoint regulator, let \(N_{\rm grid}\) denote the number of radial
intervals and
\begin{equation}
 n_g=N_{\rm grid}-1
 \label{eq:prd-n-interior}
\end{equation}
the number of interior integration variables for each field.  If
\(\mathbb O_{i,n_g}\) denotes the \(2n_g\times2n_g\) matrix obtained by
discretizing the differential operator in the proper radial coordinate,
the matrix that appears in the finite-dimensional quadratic action is
not \(\mathbb O_{i,n_g}\) itself but
\begin{equation}
 \mathbb K_{i,n_g}=h_i\,\mathbb O_{i,n_g},
 \qquad
 h_i=\frac{L_i}{N_{\rm grid}}.
 \label{eq:prd-action-second-variation-vs-operator}
\end{equation}
This relation holds for the diagonal derivative terms, the
first-derivative mixing term, and the local potential terms when the same
midpoint prescription is used throughout.

The normalized time-sliced measure for the two canonical variables
\(g\) and \(s\) is
\begin{equation}
 {\cal D}\mu_{i,n_g}
 \propto
 h_i^{-N_{\rm grid}}
 \prod_{j=1}^{n_g}dg_j\,ds_j .
 \label{eq:prd-measure-jacobian}
\end{equation}
For each background \(i={\rm B},{\rm FV}\), let
\(Z_{i,N_{\rm grid}}^{(2)}\) denote the magnitude of the regulated
\(O(4)\)-symmetric contribution to the path integral in the Gaussian approximation,
evaluated with the measure prescription above.  Here \(N_{\rm grid}\) is the number of radial intervals
and \(n_g=N_{\rm grid}-1\) is the number of interior variables for each field.
Consequently,
\begin{equation}
\begin{aligned}
 Z_{i,N_{\rm grid}}^{(2)}
 &\propto
 h_i^{-N_{\rm grid}}
 \left|\det\mathbb K_{i,n_g}\right|^{-1/2}
 \\
 &=
 h_i^{-(N_{\rm grid}+n_g)}
 \left|\det\mathbb O_{i,n_g}\right|^{-1/2},
 \qquad i={\rm B},{\rm FV}.
\end{aligned}
 \label{eq:prd-regulated-Gaussian}
\end{equation}
The absolute value refers to the determinant magnitude; the contour phases
associated with the negative directions are treated separately.  For the
bounce and false vacuum we use the same \(N_{\rm grid}\) and the same measure
prescription.  Since \(h_i=L_i/N_{\rm grid}\), their ratio is therefore
\begin{equation}
 \left|
 \frac{Z_{{\rm B},N_{\rm grid}}^{(2)}}
      {Z_{{\rm FV},N_{\rm grid}}^{(2)}}
 \right|
 =
 \left(\frac{L_{\rm FV}}{L_{\rm B}}\right)^{N_{\rm grid}+n_g}
 \left|
 \frac{\det\mathbb O_{{\rm B},n_g}}
      {\det\mathbb O_{{\rm FV},n_g}}
 \right|^{-1/2}.
 \label{eq:prd-regulated-Z-ratio}
\end{equation}
We define the logarithm of this measure-normalized relative Gaussian
contribution by
\begin{equation}
 \Delta_{N_{\rm grid}}^{(\ell=0)}
 \equiv
 \log\left|
 \frac{Z_{{\rm B},N_{\rm grid}}^{(2)}}
      {Z_{{\rm FV},N_{\rm grid}}^{(2)}}
 \right| .
 \label{eq:prd-DeltaN-definition}
\end{equation}
Using \(N_{\rm grid}=n_g+1\), this becomes
\begin{equation}
  \Delta_{N_{\rm grid}}^{(\ell=0)}
 =
 -\frac12
 \log
 \left|
 \frac{\det\mathbb O_{{\rm B},n_g}}
      {\det\mathbb O_{{\rm FV},n_g}}
 \right|
 +(2n_g+1)\log\frac{L_{\rm FV}}{L_{\rm B}} .
 \label{eq:prd-DeltaN}
\end{equation}
The apparently large length-dependent terms in the bare operator
determinant are therefore not physical ultraviolet contributions by
themselves: they must be combined with the numerical-integration factor in the
second-variation matrix of the action and with the normalization of the functional measure.

\begin{figure}[t]
 \centering
 \includegraphics[width=0.88\columnwidth]{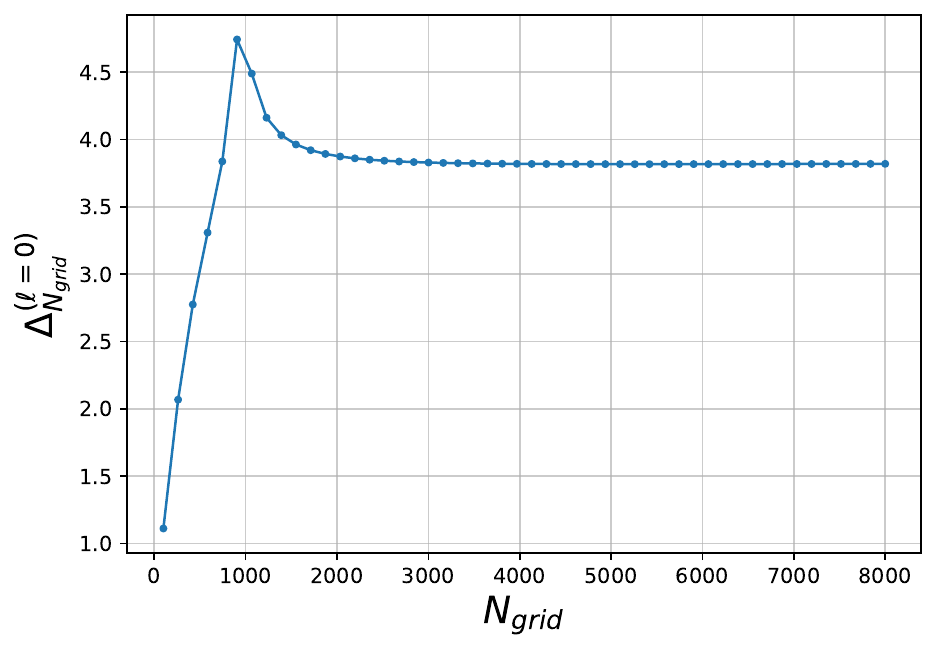}
 \caption{Convergence of the \(O(4)\)-symmetric
 relative determinant, \(\Delta_{N_{\rm grid}}^{(\ell=0)}\), defined in
 Eq.~\eqref{eq:prd-DeltaN}, for the representative value
 \(V_0=0.01M_{\rm Pl}^4\) and the parameters in
 Eq.~\eqref{eq:prd-quartic-parameters}.  Here \(N_{\rm grid}\) is the number
 of radial intervals.}
 \label{fig:prd-determinant-convergence}
\end{figure}

The continuum quantity is
\begin{equation}
 \Delta_{\ell=0}
 =
 \lim_{N_{\rm grid}\to\infty}\Delta_{N_{\rm grid}}^{(\ell=0)} .
 \label{eq:prd-Delta-continuum}
\end{equation}
We first verify the numerical convergence of the relative determinant at
the representative value \(V_0=0.01M_{\rm Pl}^4\), with the remaining
parameters fixed as in Eq.~\eqref{eq:prd-quartic-parameters}.
Figure~\ref{fig:prd-determinant-convergence} shows the convergence of
\(\Delta_{N_{\rm grid}}^{(\ell=0)}\) with increasing \(N_{\rm grid}\).

\begin{figure}[t]
 \centering
 \includegraphics[width=0.88\columnwidth]{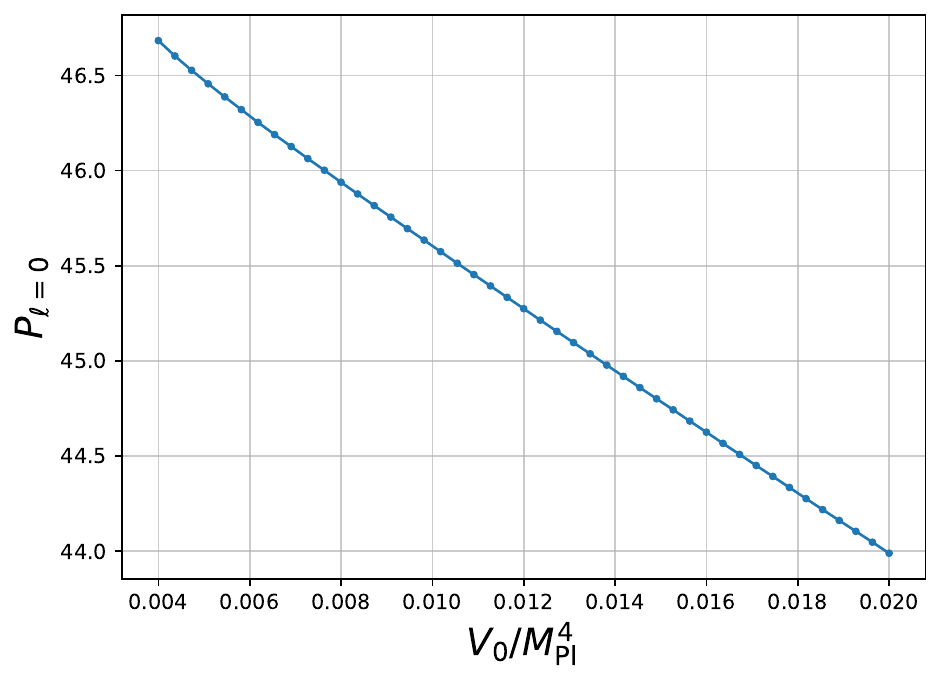}
 \caption{Relative \(O(4)\)-symmetric determinant factor
 \(P_{\ell=0}=\exp(\Delta_{\ell=0})\) as a function of the de Sitter
 vacuum energy \(V_0\), with the parameters in
 Eq.~\eqref{eq:prd-quartic-parameters} held fixed.  All determinant
 calculations use \(N_{\rm grid}=8000\) and the measure normalization of
 Eq.~\eqref{eq:prd-DeltaN}.}
 \label{fig:prd-V0-scan}
\end{figure}

We then keep the parameters in Eq.~\eqref{eq:prd-quartic-parameters}
fixed and vary the additive de Sitter vacuum energy over
\begin{equation}
 0.004\leq\frac{V_0}{M_{\rm Pl}^4}\leq0.020.
\end{equation}
For each value of \(V_0\), we construct the corresponding compact CDL
bounce and false-vacuum background and evaluate the relative
\(O(4)\)-symmetric determinant at \(N_{\rm grid}=8000\), using the same
midpoint discretization and measure normalization described above.
Figure~\ref{fig:prd-V0-scan} shows the resulting \(P_{\ell=0}\) as a
function of \(V_0\).  The determinant factor varies smoothly throughout
the interval.

\subsection{The relative tunneling phase}

The contour prescription has already been defined at finite regulator in
Sec.~\ref{sec:common-formulation}: the common ultraviolet gravitational
negative sector is assigned the same local steepest-descent continuation
on the bounce and false vacuum, so its regulator-dependent phase cancels
in the normalized ratio. {A closely related cancellation was obtained in the small-backreaction
analysis of Ref.~\cite{Ivo2025}, where it follows from the factorization
of the gravity--matter path integral.  Here we examine the corresponding
relative cancellation directly at the level of the coupled fluctuation
spectrum, without assuming such a factorization.}  The matched spectrum of
Sec.~\ref{sec:numerical-comparison} leaves the finite relative index
\begin{equation}
 \Delta n_-^{(\ell=0)}
 =
 n_-^{\rm B}-n_-^{\rm FV}
 =1 .
 \label{eq:prd-l0-relative-count}
\end{equation}
Thus only the additional low-frequency bounce direction contributes a
relative tunneling phase in the $\ell=0$ sector.  Within the same local
Gaussian contour prescription, and up to the orientation convention for
the steepest-descent contour, this phase is $i$. The single-bounce contribution normalized by the false-vacuum saddle is
therefore schematically
\begin{equation}
 \left.
 \frac{Z_{\rm B}}{Z_{\rm FV}}
 \right|_{\ell=0}
 \sim
 \pm\frac{i}{2}\,
 P_{\ell=0}e^{-B}.
 \label{eq:prd-l0-Zratio}
\end{equation}
The factor $1/2$ is the standard factor associated with the tunneling
saddle.  The magnitude is defined by
$P_{\ell=0}=\exp(\Delta_{\ell=0})$.  

This is not the complete four-dimensional decay rate.  The $\ell=1$
collective and gauge sector, all $\ell\ge2$ fluctuations, their ghost
contributions, and the large-$\ell$ renormalization remain to be
included.  The common formulation nevertheless separates the shared
high-frequency gravitational sector from the additional
bounce-specific tunneling direction and provides a well-defined
regulated quantity whose continuum behavior can be tested numerically.

\section{Discussion and outlook}
\label{sec:discussion}

The central result of this work is that the rapidly oscillating negative
modes of Coleman--De Luccia tunneling need not be interpreted as an
infinite set of bounce-specific instabilities.  In the reduced
description, a finite interval with $Q<0$ produces arbitrarily
short-wavelength negative modes.  In the unreduced formulation,
however, the local quadratic form remains regular across $Q=0$ and,
where the scalar profile is nontrivial, retains one positive and one
negative local direction.  The singular $1/Q$ factor appears only after
the lapse fluctuation is eliminated.

The configuration-space and spectral analyses give complementary
interpretations of this result.  Near the wall, a large scalar gradient
produces strong scalar--gravity mixing, whereas at the maximal-radius
slice of a de Sitter bounce the same negative direction becomes the
gravitational scale-factor direction.  These are two ways of exposing
one underlying gravitational negative sector.  Correspondingly, when
the bounce and false vacuum are treated with the same variables, gauge,
boundary conditions, regulator, and conformal prescription, their
high-frequency negative spectra approach one another.  The common rapid
sector is thereby separated from a single low-frequency negative
direction specific to the bounce and associated with tunneling.

{
It is useful to place this conclusion alongside the recent result of
Ref.~\cite{Ivo2025}. That work demonstrates explicitly that the
common gravitational contribution cancels from the
bounce-to-false-vacuum ratio in the small-backreaction limit.
The new element studied here is the finite-backreaction regime in
which the gravitational and scalar fluctuations are mixed and the
reduced Coleman--De Luccia problem develops the rapid negative-mode
tower. The ultraviolet eigenvalue pairing and the regulated relative
count found above provide, in the $O(4)$-symmetric sector, a spectral
counterpart to the cancellation that is manifest through factorization
in the small-backreaction limit.
}

The same common formulation yields a finite relative one-loop
contribution in the $O(4)$-symmetric sector.  We have examined a
continuous family of compact de Sitter CDL solutions by varying the
additive vacuum energy over
$0.004\leq V_0/M_{\rm Pl}^4\leq0.020$, while keeping the
scalar-barrier shape fixed.  Throughout this range, the relative
determinant varies smoothly, the relative negative-mode count remains
one, and $Q_{\min}$ remains negative.  The significance of the result
is therefore not tied to a particular parameter point, but lies in the
stability of the relative spectral and determinant structure across this
family of compact CDL backgrounds.  The rapid negative sector does not
obstruct a consistent relative determinant once the common gravitational
contribution is treated in the same way on the two backgrounds.

Several issues remain open.  Our treatment of the conformal sector is
based on a local Gaussian prescription and does not determine the full
nonlinear gravitational integration cycle.  In addition, the present calculation
is restricted to the $O(4)$-symmetric sector.  The $\ell=1$ sector is
qualitatively different from the radial problem considered here: the
scalar-type harmonic decomposition is degenerate in this sector, and
collective-coordinate directions are intertwined with diffeomorphism
gauge modes.  Its contribution to the prefactor therefore requires a
consistent treatment of the gauge-fixed measure, zero-mode Jacobians,
and the associated ghost sector rather than a straightforward extension
of the $\ell=0$ determinant calculation.  For $\ell\ge2$, the scalar,
vector, and tensor fluctuation determinants, their ghost contributions,
and the renormalization of the large-$\ell$ partial-wave sum must also
be included.  These ingredients are required for the complete
four-dimensional prefactor and are beyond the scope of the present work.

\begin{acknowledgments}
{
I am grateful to Victor Ivo for bringing Ref.~\cite{Ivo2025} to my attention and for pointing out its close connection to the relative-phase interpretation discussed in this work.
}
This work is supported by Grant-in-Aid for Scientific Research from the Ministry of Education, Culture, Sports, Science and Technology, Japan (JP24K07022). The author used ChatGPT (OpenAI, GPT-5.6) as an aid for scientific discussion, numerical exploration, code development, and manuscript preparation. The author reviewed and verified the resulting calculations and interpretations and takes full responsibility for the content.
\end{acknowledgments}

\appendix

\section{Scalar harmonics on $S^3$ and the $\ell=0$ sector}
\label{app:S3-harmonics}

The main text focuses on the $O(4)$-symmetric radial fluctuations.
To place these within the full scalar-type fluctuation problem, we
summarize the $S^3$ harmonic decomposition following
Ref.~\cite{Weinberg2014}.  In this decomposition, the $O(4)$-symmetric
fluctuations correspond to $\ell=0$, while modes with $\ell>0$ depend on
the $S^3$ coordinates and therefore do not preserve the $O(4)$ symmetry
of the background.  Let $\bar g_{ab}$ be the metric on the unit $S^3$, and let
$\bar\nabla_a$ denote the associated covariant derivative.  The scalar harmonics satisfy
\begin{equation}
 -\bar\nabla^2Y_{\ell m}
 =
 k_\ell^2Y_{\ell m},
 \qquad
 k_\ell^2=\ell(\ell+2),
 \qquad
 \ell=0,1,2,\ldots ,
 \label{eq:app-harmonic-eigenvalue}
\end{equation}
and the degeneracy at fixed $\ell$ is
\begin{equation}
 d_\ell=(\ell+1)^2 .
 \label{eq:app-harmonic-degeneracy}
\end{equation}
From the scalar harmonics $Y_{\ell m}$, we construct the corresponding vector
and traceless tensor harmonics as
\begin{align}
 Y_a^{(\ell m)}
 &\equiv \bar\nabla_aY_{\ell m},
 \label{eq:app-scalar-vector-harmonic}
 \\
 Y_{ab}^{(\ell m)}
 &\equiv
 k_\ell^{-2}
 \left(
 \bar\nabla_a\bar\nabla_b
 +\frac{k_\ell^2}{3}\bar g_{ab}
 \right)Y_{\ell m}.
 \label{eq:app-scalar-tensor-harmonic}
\end{align}
The vector harmonic is absent for $\ell=0$, while the traceless tensor
harmonic is nontrivial only for $\ell\ge2$.

For each scalar harmonic, we write the scalar-type metric perturbation as
\begin{align}
 ds^2={}&
 \left[1+2A_{\ell m}(\xi)Y_{\ell m}\right]d\xi^2
 \nonumber\\
 &+B_{\ell m}(\xi)Y_a^{(\ell m)}\,d\xi\,dz^a
 \nonumber\\
 &+\rho^2\bar g_{ab}
 \left[1+2\Psi_{\ell m}(\xi)Y_{\ell m}\right]dz^a dz^b
 \nonumber\\
 &+2\rho^2 C_{\ell m}(\xi)Y_{ab}^{(\ell m)}dz^a dz^b .
 \label{eq:app-scalar-metric-decomposition}
\end{align}
while the scalar field is expanded as
\begin{equation}
 \phi(\xi,\Omega)
 =
 \phi(\xi)+\Phi_{\ell m}(\xi)Y_{\ell m}(\Omega) .
 \label{eq:app-scalar-field-decomposition}
\end{equation}
Equation~\eqref{eq:app-scalar-metric-decomposition} is the scalar-type
decomposition corresponding to Eqs.~(4.12)--(4.13) of
Ref.~\cite{Weinberg2014}.

For $\ell=0$, the harmonic is constant on $S^3$.  Hence
$\bar\nabla_aY_{00}=0$, and the corresponding vector and traceless tensor
harmonics vanish.  The perturbation therefore reduces to three radial functions,
\begin{equation}
 {\cal N}=1+A,\qquad
 \rho\to\rho(1+\Psi),\qquad
 \phi\to\phi+\Phi ,
 \label{eq:app-l0-reduction}
\end{equation}
after absorbing the constant normalization of $Y_{00}$ into the radial
amplitudes.  This is precisely the convention used in
Eq.~\eqref{eq:prd-fluctuation-convention}.  Thus the $\ell=0$ sector is
the $O(4)$-symmetric radial fluctuation problem studied in the main text.

The $\ell=1$ sector is special.  Since $k_1^2=3$, the identity
\begin{equation}
 \bar\nabla_a\bar\nabla_bY_{1m}
 =
 -\bar g_{ab}Y_{1m}
 \label{eq:app-l1-identity}
\end{equation}
implies that the corresponding traceless tensor harmonic vanishes, so
$C_{1m}$ is absent.  The number of independent perturbation variables and the
gauge structure of this sector are therefore different from those for generic
$\ell\ge2$.  A complete
treatment of the exceptional $\ell=1$ sector and of the higher partial
waves lies outside the present radial one-loop calculation.

\section{Discretized path-integral measure and reproducibility of the relative determinant}
\label{app:measure-determinant}

This appendix gives the finite-dimensional analysis underlying the
\(O(4)\)-symmetric determinant used in Sec.~\ref{sec:l0-prefactor}.  The
purpose is to make explicit which matrix is being determinant-evaluated,
how the radial numerical integration enters the quadratic action, and how the
normalization of the functional measure removes the regulator-dependent
power of the proper length.  No additional ultraviolet subtraction is
introduced in this derivation.

In the common \(A=0\) gauge, introduce
\begin{equation}
 g=\sqrt{6}\,M_{\rm Pl}\rho^{3/2}\Psi,
 \qquad
 s=\rho^{3/2}\Phi .
 \label{eq:app-gs-def}
\end{equation}
For the \(O(4)\)-symmetric sector, the complete quadratic action used in the
numerical calculation can be written as
\begin{align}
 \frac{S^{(2)}}{2\pi^2}
 &=\int_0^L d\xi
 \left[
 -\frac12\dot g^{\,2}
 +\frac12\dot s^{\,2}
 +c\,g\dot s
 \right.
 \nonumber\\
 &\hspace{3.0cm}\left.
 +\frac12U_g g^2+Bgs+\frac12U_s s^2
 \right].
 \label{eq:app-continuum-quadratic}
\end{align}
To make the numerical construction fully reproducible, define
\begin{equation}
 H\equiv\frac{\dot\rho}{\rho},
 \qquad
 c=\sqrt{\frac32}\frac{\dot\phi}{M_{\rm Pl}} .
 \label{eq:app-H-c-def}
\end{equation}
The remaining coefficient functions are
\begin{align}
 U_g
 &=
 \frac34H^2+\frac12\dot H
 +\frac{\dot\phi^2}{2M_{\rm Pl}^2}
 +\frac{V}{M_{\rm Pl}^2},
 \label{eq:app-Ug-def}
 \\
 U_s
 &=
 \frac94H^2+\frac32\dot H+V'',
 \label{eq:app-Us-def}
 \\
 B
 &=
 \sqrt{\frac32}\frac{1}{M_{\rm Pl}}
 \left(V'-\frac32H\dot\phi\right).
 \label{eq:app-B-def}
\end{align}
Here \(V\), \(V'\), and \(V''\) are evaluated on the background
\(\phi(\xi)\).  These expressions follow directly from the second
variation of Eq.~\eqref{eq:prd-action} in the common \(A=0\) gauge and
therefore apply to both the bounce and false-vacuum backgrounds.  In
particular, on a constant false vacuum, \(\dot\phi=V'=0\), so that
\(c=B=0\).

After integration by parts this is
\begin{equation}
 \frac{S^{(2)}}{2\pi^2}
 =
 \frac12\int_0^L d\xi\,
 X^T{\cal O}X,
 \qquad
 X=
 \begin{pmatrix}g\\s\end{pmatrix},
 \label{eq:app-continuum-operator-form}
\end{equation}
with
\begin{equation}
 {\cal O}
 =
 \begin{pmatrix}
 \partial_\xi^2+U_g
 &
 c\,\partial_\xi+B
 \\
 -c\,\partial_\xi+B-\dot c
 &
 -\partial_\xi^2+U_s
 \end{pmatrix}.
 \label{eq:app-full-operator}
\end{equation}

The variables in Eq.~\eqref{eq:app-gs-def} are also natural from the
configuration-space geometry.  The scalar norm contains
\(\rho^3\Phi^2\), while the \(O(4)\)-symmetric scale-factor direction of
the DeWitt norm contains \(-6M_{\rm Pl}^2\rho^3\Psi^2\).  Thus, up to an
overall convention, the configuration-space norm takes the form
\begin{equation}
 \int_0^L d\xi\,(-dg^2+ds^2).
 \label{eq:app-config-metric}
\end{equation}
The minus sign is the gravitational conformal direction.  For the
determinant magnitude, it is treated by a steepest-descent contour, as
discussed in the main text.

Divide the interval into \(N_{\rm grid}\) equal segments,
\begin{equation}
 \xi_j=jh,
 \qquad
 h=\frac{L}{N_{\rm grid}},
 \qquad
 j=0,\ldots,N_{\rm grid},
 \label{eq:app-grid}
\end{equation}
and impose the endpoint conditions used in the numerical calculation.
There are
\begin{equation}
 n_g=N_{\rm grid}-1
 \label{eq:app-n}
\end{equation}
interior variables for each field.  Define midpoint values by
\begin{equation}
 g_{j+1/2}=\frac{g_{j+1}+g_j}{2},
 \qquad
 s_{j+1/2}=\frac{s_{j+1}+s_j}{2},
 \label{eq:app-midpoint-values}
\end{equation}
and finite differences by
\begin{equation}
 \dot g_{j+1/2}
 =
 \frac{g_{j+1}-g_j}{h},
 \qquad
 \dot s_{j+1/2}
 =
 \frac{s_{j+1}-s_j}{h}.
 \label{eq:app-midpoint-derivatives}
\end{equation}
The derivative terms in Eq.~\eqref{eq:app-continuum-quadratic} become
\begin{align}
 -\frac12\int d\xi\,\dot g^{\,2}
 &\ \to\ 
 -\frac12\sum_{j=0}^{N_{\rm grid}-1}
 \frac{(g_{j+1}-g_j)^2}{h},
 \label{eq:app-disc-gkin}
 \\
 \frac12\int d\xi\,\dot s^{\,2}
 &\ \to\ 
 \frac12\sum_{j=0}^{N_{\rm grid}-1}
 \frac{(s_{j+1}-s_j)^2}{h}.
 \label{eq:app-disc-skin}
\end{align}
The derivative mixing is
\begin{equation}
 \int d\xi\,c\,g\dot s
\ \to\ 
 \sum_{j=0}^{N_{\rm grid}-1}
 \frac{c_{j+1/2}}{2}
 (g_{j+1}+g_j)(s_{j+1}-s_j).
 \label{eq:app-disc-mixing}
\end{equation}
Notice that the explicit factor of \(h\) from the numerical integration cancels the
\(1/h\) in the first derivative.  Finally, the local terms take the form
\begin{align}
 \frac12\int d\xi\,U_g g^2
 &\ \to\ 
 \frac{h}{2}\sum_j U_{g,j+1/2}g_{j+1/2}^2,
 \\
 \int d\xi\,Bgs
 &\ \to\ 
 h\sum_j B_{j+1/2}g_{j+1/2}s_{j+1/2},
 \\
 \frac12\int d\xi\,U_s s^2
 &\ \to\ 
 \frac{h}{2}\sum_j U_{s,j+1/2}s_{j+1/2}^2.
 \label{eq:app-disc-local}
\end{align}

Collecting the \(2n_g\) interior variables into
\begin{equation}
 {\bf X}
 =
 (g_1,\ldots,g_{n_g},s_1,\ldots,s_{n_g})^T,
\end{equation}
the regulated action is
\begin{equation}
 \frac{S_{\rm lat}^{(2)}}{2\pi^2}
 =
 \frac12{\bf X}^T\mathbb K\,{\bf X}.
 \label{eq:app-lattice-action}
\end{equation}
If \(\mathbb O\) denotes the matrix obtained by discretizing the
differential operator \({\cal O}\) with the same midpoint prescription,
then the numerical-integration factor gives
\begin{equation}
  \mathbb K=h\,\mathbb O .
 \label{eq:app-K-hO}
\end{equation}
Since \(\mathbb K\) and \(\mathbb O\) are \(2n_g\times2n_g\),
\begin{equation}
 \det\mathbb K=h^{2n_g}\det\mathbb O .
 \label{eq:app-det-KO}
\end{equation}

For one
canonically normalized coordinate, the normalized time-sliced measure
has the form
\begin{equation}
 {\cal D}\mu_q
 \propto
 h^{-N_{\rm grid}/2}
 \prod_{j=1}^{n_g}dq_j.
 \label{eq:app-one-field-measure}
\end{equation}
For the two variables \(g\) and \(s\),
\begin{equation}
  {\cal D}\mu
 \propto
 h^{-N_{\rm grid}}
 \prod_{j=1}^{n_g}dg_j\,ds_j .
 \label{eq:app-two-field-measure}
\end{equation}
Combining Eqs.~\eqref{eq:app-lattice-action},
\eqref{eq:app-det-KO}, and \eqref{eq:app-two-field-measure}, the
determinant magnitude of the regulated Gaussian integral is
\begin{align}
 |Z_{N_{\rm grid}}^{(2)}|
 &\propto
 h^{-N_{\rm grid}}|\det\mathbb K|^{-1/2}
 \nonumber\\
 &=
 h^{-(N_{\rm grid}+n_g)}|\det\mathbb O|^{-1/2}.
 \label{eq:app-Z-regulated}
\end{align}

For bounce and false vacuum calculations performed with the same \(N_{\rm grid}\),
\(h_i=L_i/N_{\rm grid}\).  Therefore
\begin{align}
 \Delta_{N_{\rm grid}}
 &\equiv
 \log\left|\frac{Z_{{\rm B},N_{\rm grid}}^{(2)}}{Z_{{\rm FV},N_{\rm grid}}^{(2)}}\right|
 \nonumber\\
 &=-\frac12\log
 \left|
 \frac{\det\mathbb O_{{\rm B},n_g}}
      {\det\mathbb O_{{\rm FV},n_g}}
 \right|
 +(N_{\rm grid}+n_g)\log\frac{L_{\rm FV}}{L_{\rm B}} .
 \label{eq:app-Delta-general}
\end{align}
Using \(N_{\rm grid}=n_g+1\),
\begin{align}
 \Delta_{N_{\rm grid}}
 &=-\frac12\log
 \left|
 \frac{\det\mathbb O_{{\rm B},n_g}}
      {\det\mathbb O_{{\rm FV},n_g}}
 \right|
 \nonumber\\
 &\quad +(2n_g+1)\log\frac{L_{\rm FV}}{L_{\rm B}} .
 \label{eq:app-Delta-2n1}
\end{align}

\end{document}